\documentclass[namedreferences]{solarphysics}

\usepackage[hyperref,optionalrh]{spr-sola-addons} 
\usepackage{graphicx}        
\usepackage{color}           
\usepackage{breakurl}        

\ifx\urlurl    \undefined
  \def\urlurl#1{\href{http://#1}{\textsf{#1}}}\fi

\newcommand{\aap}{    {\it Astron. Astrophys.}}

\newcommand{\aj}{     {\it Astron. J.}} 
\newcommand{\apj}{    {\it Astrophys. J.}}
\newcommand{\apjs}{   {\it Astrophys. J. Supp. Ser.}} 
\newcommand{\apjl}{   {\it Astrophys. J. Lett.}}

\newcommand{\grl}{    {\it Geophys. Res. Lett.}}

\newcommand{\jgr}{    {\it J. Geophys. Res.}}
\newcommand{\mnras}{  {\it Mon. Not. Roy. Astron. Soc.}}

\newcommand{\prl}     { {\it Phys. Rev. Lett.}} 
\newcommand{\physscr} { {\it Phys. Scr.}} 
\newcommand{\procspie}     { {\it Proc. Soc. Photo-Opt. Instrum. Eng. (SPIE)}}  
\newcommand{\solphys}{{\it Solar Phys.}}
 
\newcommand{\ssr}{    {\it Space Sci. Rev.}} 
\chardef\us=`\_

\begin{document}

\begin{article}
\begin{opening}

\title{Irradiance Variations of the Sun and Sun-Like Stars -- Overview of Topical Collection}

\author[addressref={aff1},email={Greg.Kopp@lasp.colorado.edu}]{\inits{G}\fnm{Greg}~\lnm{Kopp}}
\author[addressref=aff2,corref,email={shapiroa@mps.mpg.de}]{\inits{A}\fnm{Alexander}~\lnm{Shapiro}}


\address[id=aff1]{Laboratory for Atmospheric and Space Physics, University of Colorado Boulder, Boulder, Colorado, USA}
\address[id=aff2]{Max-Planck-Institut für Sonnensystemforschung, Göttingen, Germany}

\runningauthor{G. Kopp, A. Shapiro}
\runningtitle{Irradiance Variations of the Sun and Sun-Like Stars}

\begin{abstract}
This topical collection summarizes recent advances in observing and modeling irradiance variations of the Sun and Sun-like stars, emphasizing the links between surface magnetic fields and the resulting solar and stellar variability. In particular, the articles composing this collection summarize recent progress in i) solar-irradiance measurements; ii) modeling of solar- and stellar-irradiance variability; and iii) understanding of the effects of such variability on Earth's climate and exoplanet environments. This topical-collection overview article gives background and more details on these aspects of variability.
\end{abstract}
\keywords{Solar Irradiance; Stellar Variability; Solar Cycle}
\end{opening}

\section{Introduction}
     \label{S-Introduction} 
The Sun provides nearly all the energy driving the Earth's climate system \citep[][]{Kren2017}, so, in addition to its intrinsic astrophysical interest,
monitoring and understanding changes in the total solar irradiance (TSI), the spatially and spectrally integrated radiative output of the Sun incident at the top of the Earth's atmosphere (and normalized to a distance of one astronomical unit from the Sun), are critical for studying climate change. The terrestrial atmospheric and climate systems respond to variations in solar radiative output on timescales from days to decades, and there is also evidence for solar influences on climate over longer timescales \citep[see, e.g.,][]{grayetal2010, Lean2010, Ineson2011, Haigh2011, Lean2017, PMIP4, CMIP6}. These responses have spectral dependencies, requiring measurements of the spectral solar irradiance (SSI) and an understanding of the Earth's atmosphere/chemistry interactions. 

The variability of solar irradiance is also of interest to stellar astronomers, who have been comparing solar variability with that of other lower main-sequence stars. In particular, there is an ongoing debate \citep[see, e.g.,][and references therein]{witzkeetal2018, Judge2020} on whether the level of solar-irradiance variability is anomalously low in comparison to the brightness variability of Sun-like stars measured during the synoptic programs of the Lowell and Fairborn observatories \citep[][]{lockwoodetal2007, halletal2009, Radicketal2018}. Furthermore, the advent of stellar photometric surveys with unprecedented precision brought about by planetary-transit searches, such as the \textit{Kepler} mission \citep{KEPLER}, escalated the interest in solar--stellar comparisons \citep[see, e.g., a review by][]{Basri_book} since understanding the physics behind solar variability helps assess stellar-brightness variations and the resulting limitations on the detectability and habitability of exoplanets.

Top-of-the-Earth's-atmosphere irradiance measurements are necessary to acquire the needed accuracies and stabilities for monitoring solar variability, as the dynamic Earth's atmosphere absorbs or scatters approximately 30\,\% of the incident net radiation and nearly all of the incident solar ultraviolet radiation. Space-borne solar-irradiance measurements commenced in the late 1970s \citep{hoytetal1992} for the TSI and select ultraviolet regions and have been uninterrupted since. Recent TSI measurements have not only extended this four-decade space-based Sun-as-a-star solar-variability data record, but have also provided improvements in accuracy, stability, and precision, helping better determine long-term trends in solar variability \citep{koppetal2014}. New SSI instruments now provide measurements through the visible and into the near infrared, sampling the dominant spectral region driving Earth's climate and atmospheric systems with daily temporal cadence.

There has also been significant progress in modeling and understanding solar-irradiance variability. Empirical and semi-empirical solar-irradiance models can now reproduce the measurements of solar-irradiance variability in great detail  \citep{TOSCA2013, MPS_AA, Coddington2019}. Currently, a new generation of irradiance models based on realistic 3D magneto-hydrodynamic simulations and comprehensive radiative-transfer codes with improved atomic and molecular data are being developed \citep[see, e.g.,][]{minSun}. These models are free from empirical adjustments and provide important constraints on historical solar-irradiance variability. Furthermore, they can be readily extended to modeling brightness variations of stars with different fundamental parameters and activity levels, providing background signals against which transiting exoplanets must be detected.


This \href{https://link.springer.com/journal/11207/topicalCollection/AC\_aa6896b3731736dca3a957b42f0e2e25/page/1}{topical collection}  includes articles spanning the observational and modeling aspects of solar and stellar variability. Many of these submissions are outcomes of three recent meetings discussing these topics:
\begin{itemize}
    \item 2018 Sun--Climate Symposium (\urlurl{lasp.colorado.edu/home/sorce/news-events/meetings/2018-scs/}; \urlurl{lasp.colorado.edu/home/sorce/news-events/meetings/2018-scs/})
    \item Cool Stars 20 Splinter ``Stellar Brightness Variations: building on the solar knowledge” (\urlurl{solve.mps.mpg.de/organize\_cool\_stars.shtml}; \urlurl{solve.mps.mpg.de/organize\_cool\_stars.shtml})
    \item IAU XXXth Focus Meeting FM9: "Solar Irradiance: Physics-Based Advances” (\urlurl{solve.mps.mpg.de/organize\_IAU.shtml}; \urlurl{solve.mps.mpg.de/organize\_IAU.shtml})
\end{itemize}

We provide here an overview of solar- and stellar-brightness variability knowledge and issues. In Section~\ref{Observations}, we summarize recent solar-irradiance measurements and historical reconstructions based on extensions via solar proxies. In Section~\ref{Models}, we describe physics-based advances to solar models that help explain the observed variabilities and allow extensions to stellar-brightness variations. Section~\ref{Stellar} discusses new measurements of stellar-brightness variabilities, their potential imposed limitations on exoplanet detection, and how they place the magnitudes and causes of solar variability in the context of stellar variability. We draw the conclusions and present our view on the progress expected in the field during the next few years in Section~\ref{Conclusions}.

\section{Solar-Variability Observations}
\label{Observations}
Solar irradiance varies on all timescales at which it has ever been observed and presumably also at longer ones. Ground-based measurements of the solar irradiance began in the 1830s and continued, with improved corrections for the effects of the Earth's atmospheric absorption and scatter, into the 1950s \citep{Abbot1958}. Instrument accuracies and the difficulties of correcting for atmospheric losses limited the ability to discern variations in the solar irradiance, or what was therefore referred to as the ``solar constant." It was not until space-based measurements began in 1978 that the variability of the Sun's radiant output was definitively detected (and the misnomer of ``solar constant" was realized). It is this space-borne measurement era with ever-improving accuracies and stabilities that forms the basis of this topical collection.

\subsection{Total Solar Irradiance}
\label{TSI}
The space-borne TSI record is shown in Figure~\ref{TSI} and includes measurements from nearly 20 different space-borne instruments. Improvements in absolute accuracy, particularly with the launch of the new \textit{Total Irradiance Monitor} (TIM) onboard NASA's \textit{SOlar Radiation and Climate Experiment} (SORCE) in 2003 \citep{rottman2005}, established the now-accepted TSI value of 1361\,W~m\textsuperscript{-2} \citep{KoppandLean2011}. Subsequent instruments, particularly \textit{Picard}/PREMOS \citep{schmutzetal2013}, TCTE/TIM, and TSIS-1/TIM, have confirmed this value. These latter instruments launched after performing calibrations or validations on the TSI Radiometer Facility \citep{Greg2007}, the world's only ground-based facility able to provide end-to-end TSI instrument comparisons against a NIST-calibrated cryogenic radiometer while operating under flight-like conditions. These improvements in absolute accuracy, which is currently at the $\approx$\,0.03\,\% level \citep{Prsa2016}, are important both for maintaining a link to the existing $>$~40-year TSI record (should a data gap occur in future measurements) and for being able to discern long-term solar-irradiance variations on multi-decadal timescales, as described by \cite{koppetal2014}. Measurement stability combined with continuity from overlapping data sets are currently relied upon for the detection of potential long-term trends in solar variability, as needed for climate studies. Although there is no on-orbit reference against which to validate the TSI instruments, current estimates for the best measurement stabilities after correcting for intrinsic-instrument degradation are $\approx$\,0.001\,\%~yr\textsuperscript{-1}.

 \begin{figure}
   \centerline{\includegraphics[width=0.75\textwidth,clip=]{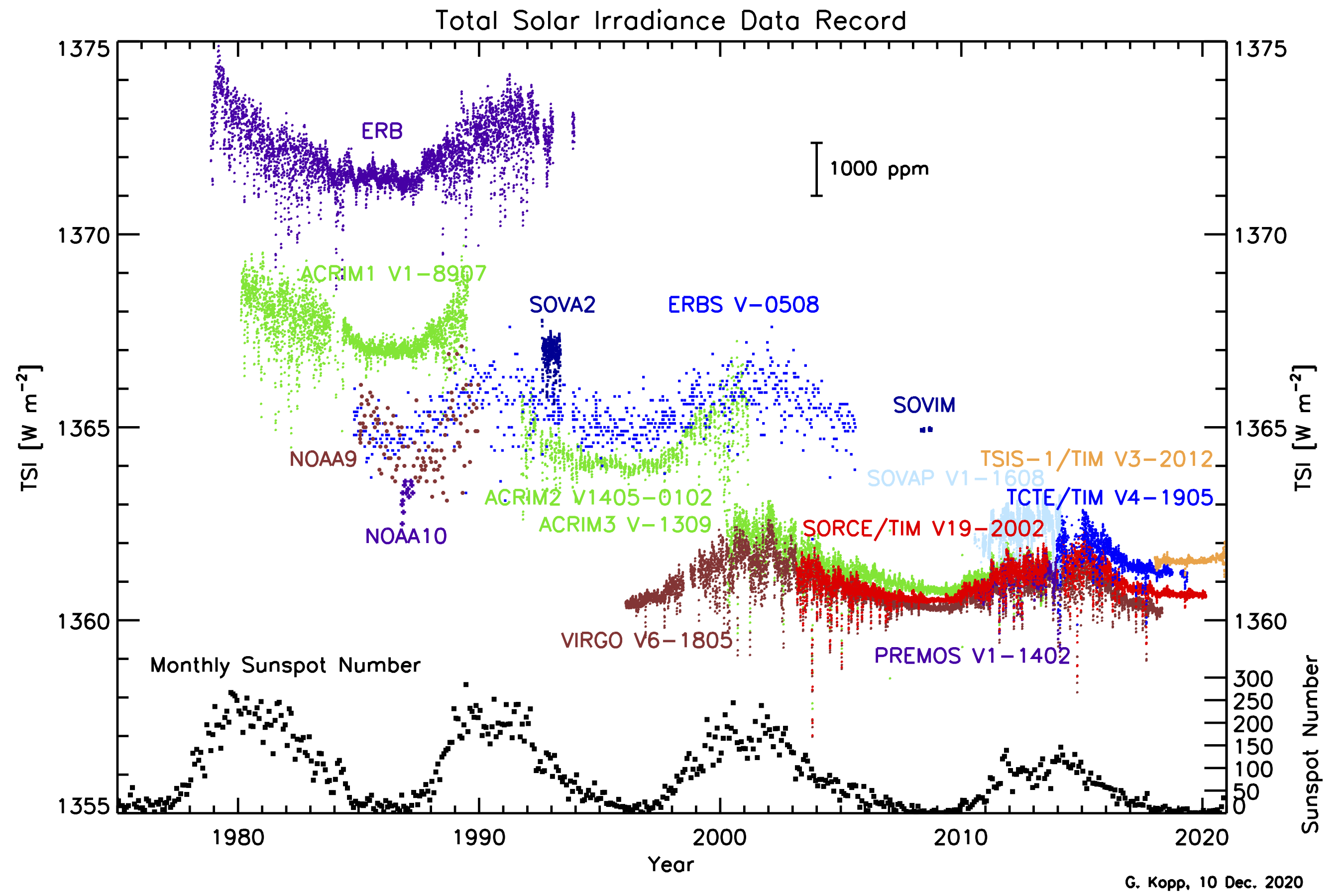}}
   \caption{The space-borne TSI record began in 1978 and has been uninterrupted since then thanks to the continuity provided by several overlapping international instruments. Note the improved consistency of the most recent instruments.}
   \label{TSIfig}
   \end{figure}

\begin{figure}
   \centerline{\includegraphics[width=0.95\textwidth,clip=]{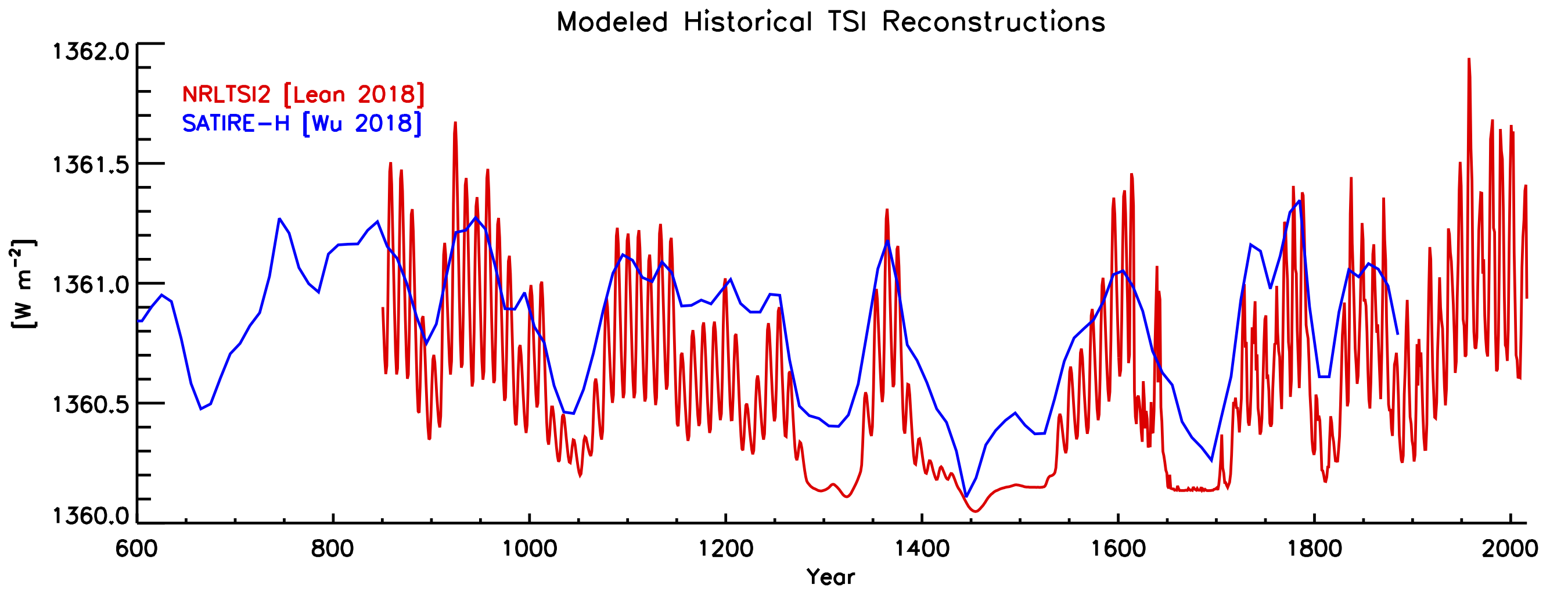}}
   \caption{TSI models enable historical reconstructions needed for climate-change studies. Two such reconstructions, the NRLTSI2 \citep{Lean2018} and SATIRE-H \citep{Wu2018} models, are shown here.}
   \label{Historical TSI}
   \end{figure}

The improved TSI-measurement accuracies, stabilities, and precision from the newest generation of instruments enable better solar-irradiance models, which are discussed in Section~\ref{Models}. Empirical and semi-empirical models reconstruct historical TSI values based on the recent-era TSI measurements and various solar proxies (see Section~\ref{Proxies}). \cite{Wu2018} and \cite{Lean2018} provide two such historical reconstructions using sunspot number and cosmogenic isotopes as proxies. These authors' reconstructions are plotted in Figure~\ref{Historical TSI}. It is mainly from such long-duration TSI records that the sensitivity of the Earth's climate to solar variability can be estimated, as the spacecraft era alone is currently too short in duration to provide the desired comparisons with long-term climate records needed for such sensitivity analyses.

\subsection{Spectral Solar Irradiance}
\label{SSI}
The processes by which the net solar energy given by the TSI measurements affect the Earth’s climate system are spectrally dependent \citep[see, e.g., reviews by][]{grayetal2010, MPS_AA}. High-altitude atmospheric absorption, troposphere/stratosphere mixing, and coupled ocean--atmospheric effects are examples of such processes. Thus, in addition to TSI, measuring the spectral distribution of the incoming solar irradiance, or the SSI, is also critical for modeling climate variability. 

A number of reference spectra, e.g. ATLAS3 \citep{thuilleretal2004} and SIRS WHI \citep{SIRS}, have been created to quantify the absolute level of the SSI from the UV to the infrared. These spectra agree reasonably well in the UV and visible spectral domains, having differences within 2\,--\,3\,\% between spectral fluxes at most of the wavelengths \citep[e.g. see][]{ATLASvsSIM}. However, there is still controversy regarding the absolute level of infrared irradiances, where deviations reaching 7\,\%  at 1700\,nm are reported \citep{IR2, IR3}. \cite{Meftah2020} recently produced a new reference spectrum from 165 to 3000\,nm based on the SSI measurements from the SOLAR/SOLSPEC instrument in an attempt to improve this situation.

To quantify the variability of the SSI, a temporally continuous record spanning nearly the entire solar spectrum is needed.  While ultraviolet SSI measurements began at about the same time as the space-borne TSI measurements, it was not until 2003 that such a record was observed from space at longer wavelengths \citep{Rottman2005_SIM}. That initial SORCE/\textit{Spectral Irradiance Monitor} (SIM) spanned the 240 to 2400\,nm spectral region, sampling $\approx$\,96\,\% of the radiant solar energy, and provided daily measurements from 2003 to 2020. A newer version of the SIM was launched in 2017 and is currently continuing these SSI measurements with improved absolute accuracies and long-term stabilities, as described by \cite{Richard2020}.


\subsection{Solar-Irradiance Composites}
\label{Composites}
\begin{figure}
   \centerline{\includegraphics[width=0.95\textwidth,clip=]{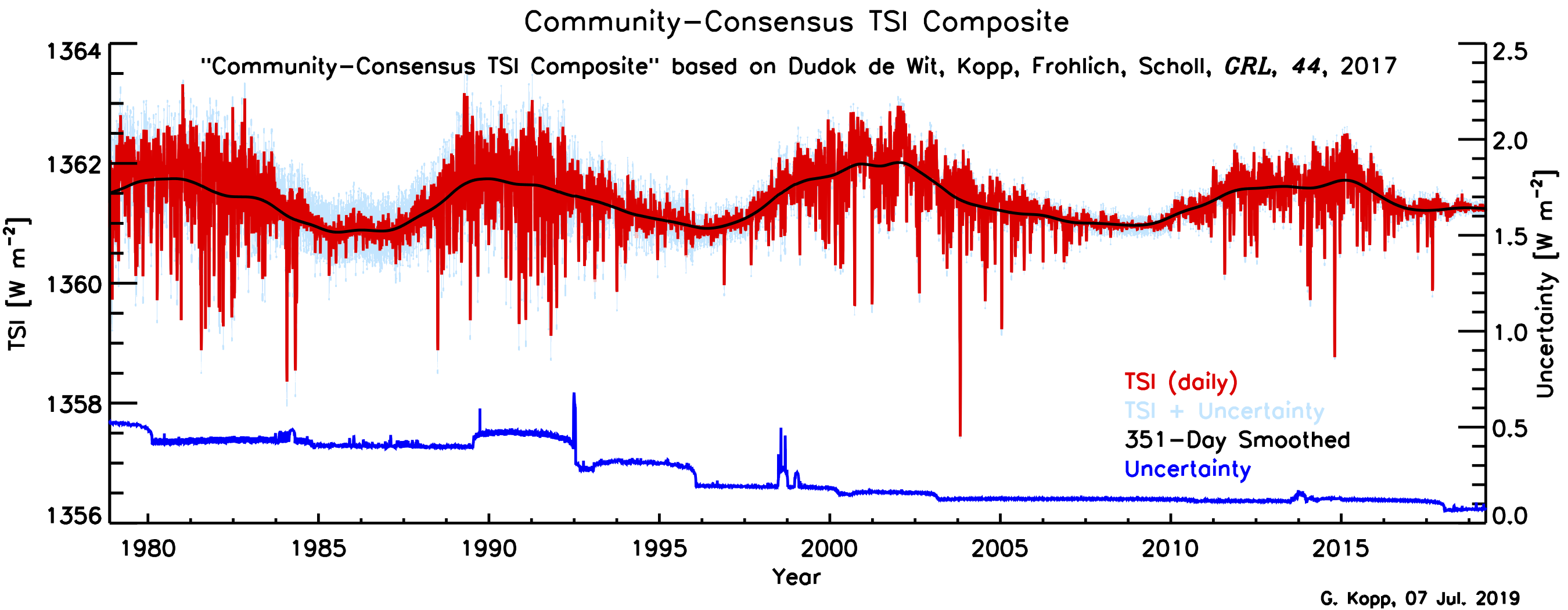}}
   \caption{Utilizing the methodology described by \cite{composite}, this ``Community-Consensus TSI Composite" provides the user community with a single time series from the multiple instruments  contributing to the four-decade-long space-borne TSI-measurement record shown in Figure~\ref{TSIfig}.}
   \label{Composite TSI}
   \end{figure}

Measurement records from individual instruments overlapping in time can be pieced together to produce a composite time series by various means. Traditionally these composites have had the following issues: they have been created by individuals (usually instrument principal investigators) and the composites produced thus reflect their biases in instrument selection; the data for any one time is provided by a single instrument rather than merging all available instruments' data, causing possible discontinuities in absolute value or slope at instrument-to-instrument transitions; and the composites generally do not include uncertainties. \cite{composite} avoid these issues with one of the most sophisticated and least-biased composite-creation methodologies by using a data-driven statistical-based method to estimate instrument uncertainties and smoothly merge scale-wise weighted data from all available instruments. An updated version of that composite, referred to as the ``Community-Consensus TSI Composite," is shown in Figure~\ref{Composite TSI}.

Combining multiple instruments can similarly extend the spectral ranges of measurements. At a relatively simple level, the three SSI instruments onboard the SORCE mission have been combined to produce an SSI time series spanning 0.1 to 40\,nm and 115 to 2416\,nm for much of the duration of that mission's lifetime. The more sophisticated SOLar Irradiance Data (SOLID) project applies statistical-based methodologies described by \cite{SOLID} to assess a broad range of SSI and proxy datasets and estimate short- and long-term uncertainties of each to create a resulting SSI composite.

\subsection{Solar Indices}
\label{Proxies}

Solar proxies are various direct or indirect indicators of solar variability. Common examples include variations in solar Fraunhofer lines, quantification of features on the solar surface, and changes affecting magnetic fields or particles within the heliosphere. Those that are affected by solar-surface magnetic activity, which are the features that cause the majority of solar-irradiance variability, can provide knowledge of such variability at times when direct measurements may not be available. Proxies can thus be useful for filling gaps in measurement data and for extending the observational records to historical times.

\cite{Choudhary_SP} link space-borne TSI measurements with ground-acquired solar proxies to describe the solar variability over the last two solar cycles. They utilize photometric indices based on Ca\,{\sc ii}\,K line and magnetograms to create an empirical proxy model of TSI measurements. Such models can help fill TSI-measurement data gaps and, if of high enough fidelity and long-term stability, they may also help discern issues with spacecraft instruments, such as abrupt changes or even long-term drifts. 

While this example relies on photometric sums that do not predate the spacecraft TSI measurements, similar models based on proxies that began before the space-borne measurement era allow extrapolations of solar-irradiance variations to historical times. The four-decade-long solar-irradiance record can be compared to longer-duration proxies that are indicative of the solar activity causing irradiance variability. Common such proxies include Mg\,{\sc ii}, Ca\,K, F$_{10.7}$, sunspot-number and -area, and the cosmogenic isotopes \textsuperscript{10}Be and \textsuperscript{14}C . These proxy records can then extend the irradiance record back in time, as described by \cite{Lean2018} and \cite{Wu2018} for the TSI. Two of these proxies, Mg\,{\sc ii} and Ca\,K, are so useful for indicating solar variability that \cite{Berrilli_SP} have created historical records of those two proxies back to 1749 based on other solar proxies. 

Since the space-borne measurement record is relatively short on climate timescales, solar proxies are currently the best means of estimating historical irradiances over the longer timescales needed for climate studies. Comparing these solar-proxy-based irradiance reconstructions with similar-duration Earth-climate records, such as historical land-surface and ocean temperatures, sea levels, ice and glacier extents, and vegetation growth, then provides a means of estimating the Earth's climate sensitivity to solar forcing. Since the Sun is a natural influence on climate, it must be discerned from other causes affecting climate for understanding, mitigating, or creating international polices regulating climate change. Historical reconstructions via solar proxies enable such a means of estimating solar-forcing sensitivities on the relevant timescales.

\section{Solar-Variability Models}
\label{Models}
\subsection{Empirical and Semi-Empirical Solar-Irradiance Models}
Magnetic fields emerging in the solar atmosphere modify the atmospheric properties and lead to the appearance of surface magnetic features, such as bright faculae and dark spots \citep[see, e.g., review by][]{Sami_B}. 
The irradiance changes caused by the surface magnetic fields depend on the exact surface distribution of the fields, which define the solar-disk coverage by dark and bright features and, consequently, the disk-integrated effect due to these features. The irradiance darkening from spots and brightening from faculae can be connected to the proxies of solar-magnetic activity, as described in Section \ref{Proxies}. For example, darkening due to spots might be calculated from the photometric sunspot index while brightening due to faculae would be based on the Mg\,{\sc ii} index. This forms the basis of empirical models of solar-irradiance variability, such as the NRLTSI and NRLSSI suite of models \citep{Lean1997, Coddington2019} and the EMPIRE model \citep{EMPIRE}. These models use available records of solar irradiance to establish empirical relationships between irradiance change at any given wavelength and proxies of solar-magnetic activity. They allow reconstructions of total and spectral solar irradiance over the entire period of the proxies. 

Semi-empirical models, such as the 1D version of SATIRE \citep{fliggeetal2000, krivovaetal2003, SATIRE}, SRPM \citep{fontenla2011, Fontenla2015, Juan2018}, and COSI \citep{shapiroetal2010, shapiro_rec}, take another approach. They employ 1D semi-empirical structures separately representing the quiet Sun (i.e. regions of the solar disk free from apparent manifestations of magnetic activity) and magnetic features to calculate the spectral contrasts of these features with respect to the quiet Sun and, consequently, their contribution  to the irradiance change \citep[see][for details]{Unruhetal1999}. The distinct semi-empirical 1D structures representing the quiet Sun and different magnetic features are constructed from their measured spectra and center-to-limb-variations \citep[see][and references therein]{Fontenla2015}. They use various methods to reconstruct the distribution of magnetic features on the solar disk for different timescales. For example, SATIRE uses direct measurements of solar-surface magnetic fields during the recent spacecraft era \citep[see][]{yeoetal2014}, sunspot records to extend back through the telescope era \citep[see][]{krivova_rec2010, Dasi2016}, and cosmogenic isotope data on longer historical timescales \citep[see][]{Wu2018}. While semi-empirical 1D structures do not rely on the records of solar-irradiance variability, SATIRE still requires TSI records to constrain a free parameter in the model \citep[see detailed discussion by][]{yeoetal2014}. A promising new generation of physics-based models is even less reliant on such constraints.

\subsection{A New Generation: Physics-Based Irradiance Models}
A major advance in irradiance modeling has become possible due to significant progress in 3D radiative magneto-hydrodynamic (MHD) simulations of flows and magnetic fields in the near-surface layers of the Sun and stars \citep[see][for a review of crucial  improvements in the treatment of radiation in such simulations]{RT_LR}.  3D radiative MHD simulations have reached a new level of realism and are now replacing semi-empirical representations of the effects of magnetic fields on the solar atmospheric radiances with direct physics-based calculations \citep[see discussion in][]{SIM_MHD}. 

\cite{Sasha_NAT} used 3D HD (hydrodynamic, i.e. without magnetic field) simulations with the MURaM code \citep{MURAM} to show that TSI variability on timescales below one day (although not including $p$-mode oscillations) is driven by the evolving cellular patterns on the solar surface caused by granulation. \cite{KL2017} created the SATIRE-3D model and reconstructed the TSI variability over the 2010\,--\,2017 period without any free parameters and, thus, without any reliance on observational records of solar-irradiance variability. They employed brightness contrasts of magnetic features calculated by \cite{norrisetal2017} based on 3D MHD simulations of the solar atmosphere performed by \cite{Beeck3} using the MURaM code. Recently, the MURaM simulations accounting for the effect of small-scale dynamo became available \citep[see, e.g.,][and references therein]{Remple2018, Rempel2020}. Utilising these simulations, \cite{minSun} found that the change of the TSI between 2019 and Maunder minimum could not be larger than $2 \pm 0.7$\,W\,m$^{-2}$. 

Another important improvement is the development of radiative-transfer codes capable of accurately accounting for deviations from Local Thermodynamic Equilibrium (LTE). Deviations from LTE are crucial in accurately modeling solar UV variability, which is especially important for climate studies due to the effects these short wavelengths have on the Earth's upper atmosphere. \cite{Rutten_SP} and \cite{Criscuoli_SP} provide recent reviews and inter-comparisons of the available non-LTE codes. While these accurate non-LTE codes can be used for calculating spectra emergent from the 3D-MHD cubes, a self-consistent treatment of non-LTE in 3D MHD simulations is still not feasible \citep[so that various simplifications are used for calculating solar chromosphere, see][]{RT_LR}.


Irradiance models that rely on LTE-only calculations of quiet and active solar-surface regions (e.g. SATIRE) require empirical corrections in the UV spectral regions, where non-LTE effects are most significant \citep[see][for a detailed discussion of the corrections implemented in SATIRE]{krivova_UV, yeoetal2014}. Recently, \cite{Rinat_2019} re-calculated the spectra utilized in SATIRE using the non-LTE radiative transfer code NESSY \citep{shapiroetal2010,Rinat_2017}, circumventing the need for this empirical correction. The UV variability found by these authors using their non-LTE modeling was very similar to that calculated by \cite{yeoetal2014} using LTE-only models plus their empirical correction.

A big hurdle in modeling solar-irradiance variability is incomplete data on atomic and molecular lines in the solar spectrum. These lines dominate the opacity in the UV and visible spectral domains and thus largely determine the amplitude of the solar-irradiance variability in these spectral regions \citep{shapiroetal2015}. The current line lists used in radiative-transfer codes include more than 100 million Fraunhofer lines \citep{kurucz2005} but nevertheless underestimate the opacity in the UV. Consequently, various adjustments must be applied to the line opacities in the UV, and these can affect calculations of the irradiance variability \citep[see detailed discussion in][]{Criscuoli_opacity}. Fortunately, new atomic  and molecular data allowing more reliable computations of the opacities in the solar atmosphere are gradually becoming available. Recent advances in laboratory astrophysics \citep[see][]{Ryabchikova_SP} and in consolidating the available data \citep[e.g. a major upgrade of the Vienna atomic line database,][]{VALD_up} will allow yet further improvements to calculations of solar-irradiance variability without having to rely on such ad-hoc opacity adjustments.

\section{Brightness Variations of Sun-like Stars}
\label{Stellar}
The recent goldmine of space-based photometric-quality stellar-variability data brought by the advent of planetary-hunting missions such as CoRoT
\citep[\textit{Convection, Rotation and planetary Transits:}][]{COROT}, {\it Kepler} \citep{KEPLER}, and {\it Transiting Exoplanet Survey Satellite} \citep[TESS:][]{TESS} triggered unprecedented advances in our knowledge of stellar variability across a broad range of stellar types by providing measured brightness variations from hundreds of thousands of stars. From the observed light curves, crucial information regarding stellar rotation periods and inclinations, activity levels and cycles, and even stellar butterfly diagrams can be inferred. 

These observations in turn led to a corresponding increase in the fidelity of stellar-brightness variability models. Such brightness-variability models are needed for purely stellar studies as well as for more reliable detection and characterization of transiting exoplanets.

The development of models of stellar-brightness variations is strongly facilitated by the recent improvements in modeling and understanding the solar-irradiance variability described in Section \ref{Models}, and many of these stellar models rely on extending the solar paradigm to Sun-like stars. Several studies \citep[e.g.][]{Lagrange2010,  Meunier2013, Meunier2015, Borgniet2015} apply this approach to study the impact of stellar activity on planetary detection with photometry, radial velocity measurements, and astrometry. Other recent studies investigated the impact of a hypothetical change of solar fundamental parameters and rotation-axis inclination on stellar variability \citep{witzkeetal2018,  Witzkeetal2020, Nemec2020}. 

One of the current limitations in extending solar models to other stars is the very limited knowledge of the properties of stellar magnetic features. The extension of the 3D radiative MHD simulations from the Sun to other main-sequence stars \citep[see, e.g.,][]{Beeck3, Salhabetal2018} with the recent inclusion of starspots \citep{Mayukh_spots} has the potential to provide such missing information and lead to new breakthroughs in modeling stellar-brightness variations.

The new stellar photometric data also led to further insights in understanding how typical the Sun is as an active star. \cite{basrietal2013} compared the TSI variability to photometric variabilities shown by G dwarfs in the {\it Kepler} field, concluding that the Sun and G dwarfs show similar levels of  variability. As a result of the very limited information on the rotation periods of {\it Kepler} stars at the time of that publication, the sample of G dwarfs utilized was not constrained to near-solar rotators, but also included stars rotating significantly faster and slower than the Sun. Since rotation period directly affects the action of the stellar dynamo and, consequently, stellar magnetic activity \citep[see, e.g.,][and references therein]{LR_Reiners}, the next step in solar-stellar comparison was constraining the stellar sample to near-solar rotators. This became possible after the rotation periods of many thousands of \textit{Kepler} stars were determined from the observed periodicities in the stellar light curves \citep[see, e.g.,][]{Timo2013, McQuillan2013}. 
Another important element needed for accurate solar--stellar comparisons has recently been brought about by the {\it Gaia} mission \citep{Gaia}, which allowed accurate characterizations of stars and, in particular, the removal of evolved stars from the comparison sample. 

Building on all of these advances, \cite{Timo2020} combined data from the {\it Kepler} and {\it Gaia} missions to show that the Sun appears to be on average five times less variable than stars having known near-solar fundamental parameters and rotation periods. 
However, these authors found that the Sun's brightness-variability level is typical for stars with near-solar fundamental parameters but whose light curves do not show a clear periodicity, meaning that they are classified as having {\it unknown} rotation periods. Interestingly, the relatively short decay time of sunspots (which very rarely survive one solar rotation) and the opposing brightness variations by faculae on the Sun \citep{Witzkeetal2020} imply that the solar rotation period would be very difficult to detect from photometric time series of the Sun \citep{Lanza_Shkolnik2014, Aigrain2015, Eliana1}, so that it would similarly be classified as having an undetected rotation period. Consequently, the results of \cite{Timo2020} suggest that the Sun would be considered a normal star in a {\it Kepler} sample of stars with near-solar fundamental parameters but unknown rotation periods. 

The existence of stars basically identical to the Sun but having significantly higher and regular variability might imply that the Sun could transition into and out of these variability levels with time. 
An alternative explanation of the high-variability stars might come from the hypothesis of \cite{Metcalfe2016}  that the solar dynamo is in a transition state between high- and low-activity regimes on a stellar-evolutionary path. According to these authors' explanation, the highly variable stars with known rotation periods (and, consequently, regular light curves) did not yet start this transition, while the Sun and other low-variability stars with non-periodic light curves have started or completed this transition. It is presently not possible to distinguish between these two scenarios (i.e. whether the Sun is occasionally passing through intermittent epochs of high variability or whether it has already evolved from highly variable to inactive). Both scenarios have different consequences for understanding the Sun's magnetic future and the resulting solar-terrestrial connection \citep[see, e.g., discussion in][]{Santos466}, giving importance to more such solar-stellar studies.  
All in all,  the stellar data provide glimpses of solar variability in the past \citep[e.g.][]{Anna2020} and future \citep[e.g.][]{Santos466}, and ``shed light" on how Sun-like the present-day Sun is.

\section{Conclusions and Future Outlook} 
\label{Conclusions}
With this overview and the associated topical collection, we describe many recent improvements to total and spectral solar-irradiance observations and models, giving better insight into the physics causing solar-irradiance variability. Higher-accuracy measurements are revealing subtleties in these irradiances that are improving knowledge of both short- and long-term changes. Those, in turn, provide tighter constraints on solar-irradiance models and the physics-based processes behind solar variability. Better historical records of solar proxies enable solar-variability reconstructions over the long time frames needed for Earth-climate and solar-variability studies. The latter, in particular, allow comparisons with stellar-brightness variations from recent exoplanet-finding missions to help place the Sun in context of thousands of similar stars, giving insight into its evolutionary path.

In addition to these recent improvements, the imminent future holds promise for many more. New flagship spacecraft missions, particularly the {\it Solar Orbiter} (SO), are acquiring measurements of the Sun giving further insights into the solar-surface magnetic structures that cause irradiance variability, including the first out-of-ecliptic magnetograms \citep{Solar_Orbiter}. This mission, launched in early 2020, will address the physics of solar effects on the heliosphere and solar-activity changes with time. Of the four primary foci of the {\it Solar Orbiter}, the one most relevant to irradiance variability is the understanding of the solar dynamo. The mission’s magnetograph, SO/{\it Polarimetric and Helioseismic Imager} \citep[PHI:][]{PHI}, will measure surface flows, advecting solar magnetic fields, the meridional flow, and solar differential rotation, helping determine the magnetic-flux transport near the poles from the mission's out-of-ecliptic vantage points. These studies are important for improving understandings of turnovers in the Sun’s activity cycle as well as for viewing the Sun as we do stars seen at higher inclination angles.

Upcoming high-resolution data from the ground-based \textit{Daniel K. Inouye Solar Telescope}  \citep[DKIST:][]{DKIST} will provide much-needed insight into the physics of magnetic-field topology, formation, and evolution. The DKIST Critical Science Plan \citep[CSP:][]{DKIST_CSP} includes several regular and synoptic programs aimed at improving our understanding of the evolution of the solar-surface magnetic field over the solar cycle and how the interaction of magnetic features with convective plasma at spatial scales unresolved by previous instrumentation affects the solar luminosity (see in particular Section 3.3 of the CSP). Synergistic observations with DKIST,  {\em Interface Region Imaging Spectrograph} (IRIS), and the {\textit Atacama Large Millimeter/submillimeter Array} (ALMA) are expected to play a crucial role for the validation and improvement of solar (MHD) models, especially those aimed at reproducing the chromospheric regions and the radiative transfer of spectral regions sampling the higher layers of the solar atmosphere.

A great deal of progress is also expected in understanding brightness variations of stars other than the Sun. While data obtained by the {\it Kepler} space telescope fully revolutionised the field of stellar activity and brightness variability, even more high-quality data are expected from the current and future planetary-hunting missions.  TESS \citep{TESS} and {\em CHaracterising ExOPlanet Satellite} \citep[CHEOPS:][]{CHEOPS} continue their scientific operations and the launch of the next flagship mission, {\it PLAnetary Transits and Oscillations of stars} \citep[PLATO:][]{PLATO}, which will measure brightness variations in up to a million stars, is planned for 2026. The new data will push forward our understanding of stellar activity as well as solar-stellar comparison studies. The later will also be greatly facilitated by data from the {\em Gaia} mission, allowing better characterizations of stars \citep[see, e.g.,][]{Bailer-Jones2020}.

These observational improvements proceed in parallel with the development of new models.  
For example, refined flux-transport models, such as the advective flux transport (AFT) model \citep{Upton, Jiang_rev}, now more realistically simulate the evolution of the surface magnetic field over the solar cycle. This allows reconstructing the evolution of the solar-surface magnetic field and irradiance over solar-rotational to multi-century timescales, which is crucial to understanding the pre-anthropogenic solar contributions to climate change from which natural sensitivities of climate can be determined. Using as inputs the recent changes to the historical sunspot record \citep{Clette2016}, this new AFT model is expected to provide better historical reconstructions of the solar-surface activity that causes irradiance variability, using updated models and data to provide an improvement to the similar flux-transport-based reconstruction given by \cite{Wangetal2005}. The applications of the flux-transport model are not limited to the Sun and they are currently being extended for modeling the evolution of surface magnetic flux on stars more active than the Sun \citep{Emre2018}. 

An improved understanding of the interactions between magnetic fields and matter in stellar atmospheres brought about by MHD simulations constitutes another important area of progress in modeling solar and stellar brightness variations.
The MHD simulations of the solar atmosphere have already been used for reconstructing TSI variability \citep{KL2017} as well as for constraining its secular component \citep{Rempel2020,minSun}.  In the absence of any direct observations of stellar magnetic features there have been a number of attempts to modify solar semi-empirical models and apply them for modeling brightness variations of solar-like stars \citep[see, e.g.,][]{witzkeetal2018,Witzkeetal2020}.
The recent MHD simulations of magnetic features on stars of different spectral classes \citep[see., e.g.,][]{Beeck3, Salhabetal2018, Mayukh_spots} will soon allow developing more realistic models of stellar variability. An extension of these simulations to stars with different values of metallicity and surface gravity will allow modeling stellar brightness variations for a broad class of solar-like stars.

{\footnotesize\paragraph*{Acknowledgements}
The authors wish to thank the helpful comments from the reviewer as well as those from the \textit{Solar Physics} staff. This work was supported by the NASA's SORCE (NAS5-97045) and SIST (NNX15AI51G) programs. AS acknowledges funding from the European Research Council under the European Union Horizon 2020 research and innovation programme (grant agreement No. 715947).
}

{\footnotesize\paragraph*{Disclosure of Potential Conflicts of Interest}
The authors declare that they have no conflicts of interest.
}

\bibliographystyle{spr-mp-sola_JL}


\end{article} 

\end{document}